\documentclass[12pt]{article}
\usepackage{graphicx}
\usepackage{pgfplots}

\usepackage[colorlinks=true,citecolor=blue]{hyperref}

\usepackage{natbib}
\usepackage{graphicx}
\usepackage{amsfonts}
\usepackage{amsmath}
\usepackage{amssymb}
\usepackage{url}
\usepackage{fancyhdr}
\usepackage{indentfirst}
\usepackage{enumerate}
\usepackage{titlesec}
\usepackage{amsthm}
\usepackage{dsfont}
\usepackage[misc]{ifsym}
\usepackage{enumitem}
\usepackage{fixltx2e}
\usepackage[pagewise]{lineno}

\pgfplotsset{compat = newest}

\usepackage{extarrows}
\theoremstyle{plain}

\newtheorem{theorem}{Theorem}
\newtheorem{lemma}[theorem]{Lemma}
\newtheorem{proposition}[theorem]{Proposition}

\theoremstyle{remark}

\theoremstyle{definition}

\newtheorem{example}[theorem]{Example}

\newtheorem{definition}[theorem]{Definition}
\newtheorem{assumption}[theorem]{Assumption}

\theoremstyle{definition}

\newcommand{\E}{\mathbb{E}}
\newcommand{\R}{\mathbb{R}}
\newcommand{\F}{\mathcal{F}}
\renewcommand{\d}{\mathrm{d}}

\newcommand{\elt}{\!\in\!}
\newcommand{\ey}{\emptyset}

 \newcommand{\ep}{~$\Box$\vspace{0.5\baselineskip}}
 
 \newcommand{\iy}{\infty}
 
 \newcommand{\Lra}{\Leftrightarrow}
 
 \newcommand{\noi}{\noindent}
 \newcommand{\pp}{\succcurlyeq}
 \newcommand{\rp}{\preccurlyeq}
 \newcommand{\spp}{\succ}
 \newcommand{\srp}{\prec}
 \newcommand{\ra}{\rightarrow}
 \newcommand{\Ra}{\Rightarrow}
 \newcommand{\sbs}{\subset}
 \newcommand{\sps}{\supset}

 \newcommand{\SC}{{\cal C}}
 
 \newcommand{\SF}{{\cal F}}

\newcommand{\ga}{\alpha}
\newcommand{\xb}{\beta}

\newcommand{\xee}{\varepsilon}

\newcommand{\gl}{\lambda}

\newcommand{\gs}{\sigma}

\newcommand{\gt}{\tau}
\newcommand{\gw}{\omega}
\newcommand{\gww}{\Omega}

\usepackage[onehalfspacing]{setspace}

\usepackage{bm}
\usepackage{tikz-qtree}
\usepackage{tikz}

\title{Anticomonotonicity for Preference Axioms:
 The Natural Counterpart to Comonotonicity}
\author{Giulio Principi\thanks{Department of Economics, New York University, USA.  \Letter~{\url{gp2187@nyu.edu}}}
\and Peter P.\ Wakker\thanks{Erasmus School of Economics, Erasmus University Rotterdam, The Netherlands. \Letter~{\url{wakker@ese.eur.nl}}}
\and Ruodu Wang\thanks{Department of Statistics and Actuarial Science, University of Waterloo,  Canada. \Letter~{\url{wang@uwaterloo.ca}}}
}
\date{\today} 

\begin{document}

	\maketitle
\thispagestyle{empty}

\begin{abstract}
Comonotonicity (``same variation'') of random variables minimizes hedging possibilities and has been widely used, e.g., in Gilboa and Schmeidler's ambiguity models. This paper investigates anticomonotonicity (``opposite variation''; abbreviated ``AC''), the natural counterpart to comonotonicity. It minimizes leveraging rather than hedging possibilities. Surprisingly, AC restrictions of several traditional axioms do not give new models. Instead, they strengthen the foundations of existing classical models: (a) linear functionals through Cauchy's equation; (b)
 Anscombe-Aumann expected utility; (c)
as-if-risk-neutral pricing through
 no-arbitrage; (d) de Finetti's bookmaking foundation of Bayesianism using subjective probabilities;
(e) risk aversion in Savage's subjective expected utility.
 In each case, our generalizations show where the critical tests of classical axioms lie: in the AC cases (maximal hedges). We next present examples where AC restrictions do essentially weaken existing axioms, and do provide new properties and new models.
\\  ~  \\
\textbf{Keywords}:  Comonotonicity,  bookmaking, hedging, subjective expected utility, ambiguity aversion \\ 
\textbf{JEL Classification}:  D81, C60, C02
 \end{abstract}

\newpage


\raggedright
\parindent 5ex

\section{Introduction}

\noi Comonotonicity is widely used in mathematics (\citealp{HLP34}, Theorem 236) and in many applied fields, including decision theory.\footnote{For comonotonicity in decision theory, see
\cite{GR16}. Further examples include fuzzy set theory (\citealp{GMS00}), insurance
(\citealp{DDGKV02}), labor market equilibria (\citealp{CES17}), multiattribute utility theory
(\citealp{EGH12}), optimal transport (\citealp{GA16}), risk allocations (\citealp{R13}), risk attitudes (\citealp{Y69}, p.\ 328),
risk measures (\citealp{FS16}), time preference (\citealp{BF23}), and welfare theory (\citealp{E04}).}
\cite{PW15} provided a survey. Comonotonicity was the main tool in \citeauthor{G87}'s (\citeyear{G87}) and \citeauthor{S89}'s (\citeyear{S89}) famous ambiguity models. Two variables are comonotonic if they covary in the same direction. 
 Comonotonicity maximizes
 leveraging possibilities while minimizing hedging possibilities (\citealp{H40}).

Anticomonotonicity (AC) is a natural counterpart to comonotonicity. Two variables $X$ and $Y$ are AC if they covary in the opposite direction; i.e., if $X$ and $-Y$ are comonotonic.
AC minimizes
 leveraging possibilities while maximizing hedging possibilities. \cite{ACV21} introduced AC diversification for Choquet integrals. AC turns out to be of interest in its own right, and this paper studies it in general. We will shed new light on many classical results, and provide new models.

\cite{S89} used comonotonicity to weaken \citeauthor{AA63} (AA)'s (\citeyear{AA63}) classical independence preference condition. The latter condition characterized subjective expected utility.
 Schmeidler, thus, obtained a new preference model, Choquet expected utility. It could accommodate ambiguity aversion in \citeauthor{E61}'s (\citeyear{E61}) paradox. This result, together with \cite{GS89}, famously opened the field of decision under ambiguity, a big field today (\citeauthor{{GM16}}, \citeyear{{GM16}}; 
\citeauthor{{T15}}, \citeyear{{T15}}). Many papers have since studied the comonotonic weakening of various axioms.

It is natural to study the counterpart to the
 Gilboa-Schmeidler approach, now weakening axioms with the AC rather than the comonotonicity restriction, where leveraging is now minimized while hedging is maximized rather than the other way around. The research question then is which models result this way. We first investigated this question for the most famous result in the literature using comonotonicity: \citeauthor{S89}'s (\citeyear{S89}) generalization of
AA's subjective expected utility. The answer (Theorem \ref{Ttm.4_charact_SEU_ind}) surprised us: the AC weakening of independence does not provide any new (generalized) model at all. It still fully axiomatizes subjective expected utility, as did AA's
 full-force independence. This result can be interpreted negatively because it did not produce any new model. However, a positive interpretation is that it reinforces the classical result of AA: we generalize their result and, more specifically, show where its critical test is, namely in the AC cases. To justify or criticize their model normatively, and to verify or falsify their model empirically, only the AC cases have to be considered, and they decide.

Next, we investigated our research question for some other famous derivations of linear/affine\footnote{An
 affine function on (a subset of) a linear space is a linear function with a constant added. A linear function assigns value 0 to the origin (0). In all our theorems, representing functionals remain representing if a constant is added, so that the difference between affine and linear never matters.}
 optimization models. We considered \citeauthor{{dF31}}'s (\citeyear{{dF31}}) bookmaking. de Finetti used bookmaking to normatively defend the use of subjective probabilities and his work is considered one of the three cornerstones of Bayesianism, together with \cite{R31} and \cite{S54}. We next considered
 as-if risk-neutral pricing by a financial market. Such pricing is necessary and sufficient to avoid arbitrage possibilities. This result is a cornerstone in finance, called the fundamental theorem of asset pricing (\citeauthor{{B09}},  \citeyear{{B09}}). Finally, we considered Cauchy's functional equation, which is also widely used (\citealp{AC14}). 
In all these cases, we found that AC restrictions do not lead to new models but to generalizations and reinforcements of existing axiomatizations. For all these classical results, we more precisely identify the critical cases to be tested or investigated, i.e., when hedging is maximal. Because all these results have the same format, becoming routine from a mathematical perspective, we present formal statements of some of them in Appendix A. Demonstrating the unity (``routine'') of these results, as done in our proofs, is an additional contribution of this paper. de Finetti's bookmaking,
 AA's subjective expected utility, and
 no-arbitrage in finance are all cornerstone results in their respective fields, developed independently. We show that they all amount to the same mathematical result, and were all obtained by establishing Cauchy's equation (Theorem \ref{Ttm.1.anticomon.lin.cont}) for their certainty equivalents.

We also investigated our research question for AC restrictions of convexity, involving inequalities rather than the equalities of affinity and linearity. Under expected utility, we obtain an AC generalization of an appealing characterization of risk aversion (Theorem \ref{Ttm.7:Savage.U.conc}). Here, as before, we do not develop a new model or phenomenon but consolidate an existing result. We point out some appealing features of (our generalization of) the result, a result known to specialists but not as widely known as it deserves to be (\S\ref{Sect.concaveU}). 
We, finally, consider some ambiguity models. Here the AC restrictions do bring new phenomena, as first shown by \cite{ACV21}, whose result we generalize (Proposition \ref{Prop.8.AC=pseudo.mix.space}). Further, AC restrictions also bring new models here, more general than those without these restrictions. We provide a first example, the
 double-cautious ambiguity model (Proposition \ref{Prop_9_DoubleC}), leaving further developments to future studies.

\section{Anticomonotonic restrictions for functionals: additivity and linearity}\label{Sect:linearity}

\noi This section presents an AC generalization (Theorem \ref{Ttm.1.anticomon.lin.cont}) of the
 well-known Cauchy functional equation for several variables. Later sections will apply this generalization to decision theory and, more narrowly, to decision making under uncertainty, giving generalizations of several classic representation theorems for linear/affine functionals (Theorem \ref{Ttm.4_charact_SEU_ind} and Propositions \ref{Prop.13_ttmnoarbitrage} and \ref{Prop.15_ttmdeFinetti}). These results essentially all follow as corollaries of the Theorem in this section.

We fix $(\gww,\mathcal F)$, in which $\gww$ is a {\it state space\/} and $\mathcal F$ a
 sigma-algebra of subsets of $\gww$ called {\it events\/}. We denote by
$B(\gww,\F)$
 the set of {\it acts\/}, i.e., all bounded measurable
 real-valued functions from $\gww$ to $\R$, equipped with the
 sup-norm. Two acts $X$ and $Y$ in $B(\gww,\F)$ are \textit{comonotonic} if
 \begin{equation}\label{def.comon}
\textnormal{for~all~} \gw,\gw'\in \gww:
\left(X(\gw) -X(\gw')\right)\left(Y(\gw) -Y(\gw')\right)\geq 0.
\end{equation}
Two acts $X$ and $Y$ in $B(\gww,\F)$ are \textit{anticomonotonic} ({\it AC\/}) if $X$ and
 $-Y$ are comonotonic. Other terms used in the literature are antimonotonicity or
 counter-monotonicity. Each constant act is both comonotonic and AC with every other act.

A functional $I : B(\gww,\F)\to \R$ is \textit{additive} if

 \begin{equation}\label{def.add}
\textnormal{for~all~} X,Y:
I(X+Y)=I(X)+I(Y) .
\end{equation}
The equation is also known as Cauchy's equation (\citealp{AC66}).
{\it Monotonicity\/} holds for $I$ if $I(X) \geq I(Y)$ whenever $X(\gw) \geq Y(\gw)$ for all $\gw \in \gww$. The functional $I$ satisfies \textit{comonotonic additivity} if Eq.\ \ref{def.add} holds only for all pairs of comonotonic acts $X,Y$, while $I$ satisfies \textit{anticomonotonic additivity} (\textit{AC additivity}) if Eq.\ \ref{def.add} only holds for all pairs of AC acts $X,Y$.
Moreover, $I$ is {\it homogeneous\/} if $I(\ga X)=\ga I(X)$ for all $\ga\in \mathbb{R}$ and all $X\in B(\gww,\F)$. {\it Positive homogeneity\/} imposes the homogeneity requirement only for $\ga \geq 0$. The functional $I$ is \textit{linear} if it is additive and homogeneous.
The above definitions are extended to $I$'s defined on subdomains in the obvious manner, imposing the requirements only when all acts involved are contained in the subdomain.

\begin{theorem}[Cauchy's equation for anticomonotonicity]
\label{Ttm.1.anticomon.lin.cont}
Under (a) continuity, (b) monotonicity, or (c) finiteness of $\Omega$, AC additivity of a functional 
$I:B(\gww,\F)\to \R$
is equivalent to additivity and, furthermore, to linearity in case of (a) or (b).
\end{theorem}

All proofs are in Appendix B. A sketch of the proof of Theorem \ref{Ttm.1.anticomon.lin.cont} is as follows. 
First, AC additivity implies $I(0) = I(0+0) = 0$, and then AC additivity for $X$ and $-X$ gives $I(-X) = -I(X)$. Second, for comonotonic acts $X, Y$, AC additivity for $X+Y$ and $-Y$ gives $I(X+Y) - I(Y)  =  I(X)$ and, thus, comonotonic additivity. Third, for general $X, Y$ in a finite space, we can write each of $X$ and $Y$ as a sum of one act increasing in indexes of the state space and another act decreasing, yielding four 
 ``index-monotonic'' acts. Every pair of those four acts is either comonotonic or AC. By proper groupings in $X + Y$, the sum of these four acts, and repeated application of comonotonic and AC additivity, additivity then readily follows for general $X,Y$. Linearity for finite state spaces follows under minimal extra conditions (\citealp{AC66}). The extension of linearity to infinite state spaces first follows for simple acts and then for general acts from standard integration techniques using monotonicity or continuity. It follows that homogeneity is readily implied by additivity together with one of the other (weak) conditions that imply linearity.

In many applications, a functional $I$ is taken as primitive, for instance in production theory, price index theory, finance, or the theory of risk measures. Then the above theorem can be directly applied. The rest of this paper focuses on decision theory, where a preference relation $\pp$ is taken as primitive.

\section{Basic definitions of decision under uncertainty}\label{no.arbitrage}

\noi Besides $(\gww,\mathcal F)$ as before, we consider a set $\SC$ of {\it outcomes\/}, endowed with a binary relation $\pp$. In the preceding section, $\SC = \R$ and $\pp$ $=$ $\geq$. A {\it preference interval\/} in $\SC$ is a subset of $\SC$ that, for each pair of outcomes $x \pp z$ contained, also contains all outcomes $y$ with $x \pp y \pp z$. 

The set of {\it acts\/}, denoted $B(\gww,\F)$, contains all maps $X$ from $\gww$ to $\SC$ that are {\it bounded\/}, i.e., there exist outcomes $x, z$ such that $x \pp X(\gw) \pp z$ for all $\gw$, and {\it measurable\/}, i.e., every inverse of a preference interval is an event. Outcomes are identified with constant acts, so that $\pp$ is also a binary relation on constant acts. The {\it preference relation\/} is an extension of $\pp$ to all acts, also denoted 
$\pp$ ---no confusion will arise. In the rest of this paper, 
$\pp$ on acts is taken as primitive, and we seek to characterize phenomena through directly observable properties of $\pp$. {\it Weak ordering\/} holds if {\it completeness\/} ($X \pp Y$ or $Y \pp X$ for all acts $X,Y$) and transitivity hold. It will be implied in all our results. 
The notation $\spp$, $\sim$, $\rp$, and $\srp$ is as usual. We call $\pp$ {\it trivial\/} if $X \sim Y$ for all acts $X,Y$. 

Throughout, we assume that $\SC$ is a mixture space, which provides a convenient generalization of convex sets. Mixture spaces include money intervals in $\R$, convex sets of probability distributions, and convex sets of commodity bundles. For simplicity, readers unfamiliar with general mixture spaces may take in mind any such special case and see that all conditions below are then satisfied. We call $\SC$ a {\it mixture space\/} if it is endowed with a mixture operation. A {\it mixture operation\/} generalizes convex combinations in linear spaces. It maps $\SC \times [0,1] \times \SC$ to $\SC$ and is denoted $\ga x + (1-\ga)y$.  It is required to satisfy the following conditions:
\begin{enumerate}[label=(\roman*)]
    \item $1 x + 0 y = x$ (identity);
   \item $\ga x + (1-\ga)y = (1-\ga)y + \ga x$ (commutativity);
    \item $\ga  (\xb x +  (1-\xb) y) + (1-\ga)y = \ga \xb x + (1-\ga \xb)y$ (distributivity).
\end{enumerate}

A
 real-valued functional $I$ {\it represents\/} $\pp$, or $\pp$ {\it maximizes\/} $I$ if the preference domain is contained in the domain of $I$ and  $X \pp Y \Lra  I(X) \geq I(Y)$. A function is an {\it interval scale\/} if it is unique up to multiplication by a positive factor and addition of a constant.
{\it Subjective expected utility\/} or {\it expected utility\/}, or {\it EU\/} for short, holds if there exist a probability measure $P$ on $\mathcal F$ and a
 {\it utility function\/} $U: \SC \ra \R$ such that $\pp$ maximizes {\it expected utility\/} $\int_{\gww}U(X)\textnormal{d}P$, where this integral, called the {\it EU\/} of $X$, is assumed to be
 well-defined and finite.\footnote{In
 decision theory, there is much interest in finite additivity. We, therefore, only require finite additivity of probability measures. A necessary and sufficient condition for countable additivity can readily be added in all our results (\citealp{W93}, Proposition 4.4).}

In most of this paper, utility $U: \SC \ra \R$ will be {\it affine\/}, i.e., it satisfies:
\begin{equation}\label{def.aff.U}
\textnormal{For~all~} \ga\in \left[0,1\right] \textnormal{~and~} x,y\in \SC:
U\left(\ga x+(1-\ga)y\right)=\ga U(x)+(1-\ga)U(y) .
\end{equation}
Acts are mixed statewise and, thus, the space of acts is also a mixture space. We will follow the economic tradition of also calling affine functionals on act spaces {\it linear\/}. Thus, we say that EU is linear in probability and weights events linearly.

{\it Mixture continuity\/} holds for $\pp$ if the sets
\[\left\lbrace \ga\in \left[0,1\right]:\ga X +(1-\ga)Z \succcurlyeq Y \right\rbrace\ \textnormal{and}\ \left\lbrace \ga\in \left[0,1\right]:Y\succcurlyeq \ga X+(1-\ga)Z\right\rbrace\]
are closed for all acts $X,Y,Z$. Together with some other conditions, mixture continuity implies the existence of a certainty equivalent for each act.

We summarize:

\begin{assumption}[Structural Assumption]
\label{struct.ass.2}
 A state space $\gww$ is given with a
 sigma-algebra $\mathcal F$ and an outcome set $\SC$ that is a mixture space. The set of acts, $B(\gww,\F)$, contains all bounded measurable maps from $\gww$ to $\SC$, and $\pp$ is a binary relation on $B(\gww,\F)$.\end{assumption}

An outcome $x$ is a {\it certainty equivalent\/} ({\it CE}) of an act $X$ if $x \sim X$. In general, it does not always need to exist or be unique.
 {\it Monotonicity\/} holds for $\pp$ if $X \pp Y$ whenever $X(\gw) \pp Y(\gw)$ for all $\gw \in \gww$. The following definitions generalize previous ones. Two acts $X,Y$ are {\it comonotonic\/} if there are no states
 $\gw, \gw'$ such that
$X(\gw) \spp X(\gw')$ and $Y(\gw) \srp Y(\gw')$. Acts $X,Y$ are {\it AC\/} if there are no states $\gw, \gw'$ such that
$X(\gw) \spp X(\gw')$ and $Y(\gw) \spp Y(\gw')$.

 \section{The intuition of anticomonotonicity}\label{intuition.AC}

\noi This section presents an informal interpretation of the AC condition. The most famous appearance of comonotonicity was in \cite{S89}. He considered the special case of Structural Assumption \ref{struct.ass.2} where $\SC$ is a convex set of probability distributions over prizes, also called {\it lotteries\/}, denoted $P,Q,R$ here. A mixture $\ga P + (1-\ga)Q$ assigns probability $\ga P(E) + (1-\ga)Q(E)$ to every prize set $E$, for $0 \leq \ga \leq 1$. Thus, $\SC$ is a mixture space. Again, mixtures are transferred to acts statewise. He further assumed EU for risk (lotteries). The above setup is known as the AA setup. 
All deviations from EU over acts are then due to ambiguity, facilitating its analysis.

Under ambiguity, EU over acts is violated by interactions between events. Thus, the classical independence axiom,
\begin{equation}\label{strong.AAind}
\textnormal{~for all acts~} X,Y,C
\textnormal{~and~} 0 < \ga < 1: X \sim Y \Ra \ga X + (1-\ga)C \sim \ga Y + (1-\ga)C,
\end{equation}
which is the main axiom used by AA to axiomatize EU over all acts, is violated. 
For example, $C$'s events may ``interact'' with $Y$'s events by providing hedges reducing variations of outcomes without doing so with $X$'s events, leading to a strict preference for the safer $\ga Y + (1-\ga)C$ in Eq.\ \ref{strong.AAind} and a violation of independence. In Eq.\ \ref{strong.AAind}, $C$ denotes a ``common'' new act  that is mixed in. Because hedging occurs in mixtures, later modifications of independence in this paper will impose comonotonicity or AC restrictions on such mixtures and will concern $X,C$ and $Y,C$.

We next discuss AC, assuming ambiguity aversion. (For ambiguity seeking, similar reasonings hold with preferences reversed.) 
Comonotonicity minimizes hedging possibilities for acts $X,Y$. Schmeidler imposed independence (Eq.\ \ref{strong.AAind}) only if acts $X,C$ are comonotonic and so are $Y,C$.\footnote{Schmeidler
 also required $X,Y$ to be comonotonic, but this restriction can be omitted (as may be inferred from the yet weaker Eq.\ \ref{AAind} given later), facilitating the following intuitions.}
 Then, hedging effects are minimal and leveraging effects are maximal in both convex combinations in Eq.\ \ref{strong.AAind}, and one may conjecture that they cancel, so that Eq.\ \ref{strong.AAind} then still holds. So it does under Schmeidler's Choquet expected utility (CEU), even characterizing that theory. 

For AC, 
 leveraging is minimal and hedging is maximal. We raised the following research question: what happens if independence (Eq.\ \ref{strong.AAind}) is only imposed if both $X,C$ and $Y,C$ are AC? Our first hunch was that interaction effects, extreme again, may again balance and cancel and that the axiom will give an alternative way to axiomatize CEU.

We could not have been farther off. As it turned out, AC independence precludes any
non-neutral ambiguity attitude! AA's EU and
 full-force independence follow (Theorem \ref{Ttm.4_charact_SEU_ind} below). This result came as a surprise to us. Whereas with minimal hedging in Eq.\ \ref{strong.AAind} no ambiguity attitude is precluded, minimal leveraging leaves more space to the extent that all ambiguity attitudes under CEU are precluded.
 This result on AC may be taken as negative: AC did not bring any new model. However, a positive interpretation is that AC provides a new and stronger axiomatization of existing models, EU in this case. To justify EU normatively or descriptively, it suffices to justify independence in the AC cases. They provide the most critical cases and all other cases follow. AC independence leads to Theorem \ref{Ttm.4_charact_SEU_ind} in the following section, and to several related results discussed later.

\section{Classical linear/affine functionals}\label{class.add}

\noi In the literature using the AA setup, outcome spaces are assumed to be mixture spaces, as in this paper, and an affine utility function $U: \SC \ra \R$ is assumed. Mostly, the mixture space is assumed to be a convex set of probability distributions, with $U$ expected utility, as in the preceding section.

We now 
formally define independence. We use a weakened version because it better conveys the intuitions of conditions defined later.\footnote{This
 weakening avoids an intuitive confusion described in \S \ref{intuition.AC} because $x$ does not provide any hedge or leverage in the
 right-hand side of Eq.\ \ref{AAind} so that any canceling of such effects cannot play any role.}\label{footn.intuition}
 {\it Independence\/} holds if
\begin{equation}\label{AAind}
\begin{aligned}
\textnormal{~for all acts~} X,C,
&\textnormal{~outcomes~} x,
\textnormal{~and~} 0 < \ga < 1:
\\
&X \sim x \Ra \ga X + (1-\ga)C \sim \ga x + (1-\ga)C.
\end{aligned}
\end{equation}
In other words, any act $X$ in any mixture can be replaced by its certainty equivalent $x$.
  The condition is seemingly weaker than Eq.\ \ref{strong.AAind} in the sense of restricting general acts $Y$ in Eq.\ \ref{strong.AAind} to constant acts $x$. However, it is readily seen to be equivalent if every act has a certainty equivalent,\footnote{In
 Eq.\ \ref{strong.AAind}, replace every act with its certainty equivalent (same for $X$ and $Y$) and use transitivity.}
which holds in all results in this paper. The condition is appealing
because it justifies ``ironing out'' in mixtures (\citealp{L20}). It is convenient for comonotonic and AC generalizations because constant acts are comonotonic and AC with every other act. Schmeidler's {\it comonotonic independence\/} requires Eq.\ \ref{AAind} only if $X,C$ are comonotonic.

\begin{definition} {\it AC independence\/} holds for $\pp$ if the implication of Eq.\ \ref{AAind} is imposed only if $X$ and $C$ are AC.
\end{definition}
\noi In other words, one can replace any act in a mixture with its certainty equivalent only if the acts in the mixture are AC. The following theorem generalizes AA's classical characterization of subjective expected utility. 

\begin{theorem}\label{Ttm.4_charact_SEU_ind}
Assume Structural Assumption \ref{struct.ass.2}.
 The following three statements are equivalent.
\begin{enumerate}[label=(\roman*)]
    \item Weak ordering, monotonicity, mixture continuity, and independence hold.
    \item Weak ordering, monotonicity, mixture continuity, and AC independence hold.
    \item Subjective expected utility holds with $U$ affine.
\end{enumerate}
In (iii), $U$ is an interval scale.
\end{theorem}

We mention two other generalizations of classic representations using linear functionals to illustrate the wide applicability of AC. The results work similarly to
Theorem \ref{Ttm.4_charact_SEU_ind} and are therefore only stated verbally here for brevity. Appendix A gives complete formal statements. The first result concerns
\citeauthor{{dF31}}'s (\citeyear{{dF31}}) axiomatization of subjective expected value maximization. It rationalized subjective probabilities, which is
 well-understood nowadays but was then a conceptual breakthrough. de Finetti used a bookmaking axiom, another influential innovation: no positive linear combination of acceptable bets should lead to a sure loss. His result can be generalized like AA's EU axiomatization above, by adding an AC restriction to the axiom (Proposition \ref{Prop.15_ttmdeFinetti} in Appendix A). Here, again, the AC restriction does not bring new models but simplifies the normative task of defending the rationality of the Bayesian approach. Again, the AC cases are critical and one can focus on those. The other cases then follow.

A similar generalization can be obtained for
 as-if risk-neutral pricing in finance. Now, acts are financial assets and the functional $I$ assigns market prices to acts.
As-if risk-neutral prices are subjective expected values based on
 as-if subjective probabilities, which are the market probabilities. Such pricing is necessary to avoid arbitrage possibilities. Proposition \ref{Prop.13_ttmnoarbitrage} in Appendix A shows that arbitrage only needs to be precluded in AC cases, and already
 as-if risk-neutral pricing is implied. Again, the essence of
no-arbitrage is captured by the AC cases. And again, minimizing leveraging possibilities, which is what the AC restriction does, works differently than minimizing hedging possibilities as comonotonicity does. 

\section{Anticomonotonic convexity for concavity of utility}\label{Sect.concaveU}

\noi In the remainder of the main text, we consider AC generalizations of convexity and concavity axioms. This means that we now deal with inequalities rather than equalities and that we relax some linearities. This section maintains linearity in events/probabilities by assuming expected utility for acts on $\gww$. However, unlike all other sections, it allows for nonlinear utility: $U$ on $\SC$ need not be EU. We provide an AC axiomatization of concave utility.

Whereas mixture sets have almost exclusively been studied for affine/linear utility in the AA setup, they provide a natural domain for studying convexity and concavity of utility, the topic of this section. We thus define: $U$ is {\it concave\/} if 
 $U(\ga x + (1-\ga)y) \geq \ga U(x) + (1-\ga)U(y)$ 
for all outcomes $x$, $y$, and $0 < \ga < 1$. {\it Convexity\/} has $\leq$ instead of $\geq$. We will maintain continuity:

\begin{definition}
Utility $U$ on the mixture space $\SC$ is {\it
 mixture continuous\/} if, for all outcomes $x,y$, $U(\ga x + (1-\ga)y)$ is continuous in $\ga$.
\end{definition}
\noi The condition is implied by affinity and also by common continuity conditions on convex subsets of Euclidean spaces. Hence, it is less restrictive than most other continuity conditions.

\begin{definition}[Convexity of preference] Preferences are {\it convex\/}
if
\begin{equation}\label{def.conv}
\textnormal{~for all acts~} X,Y \textnormal{~and~} 0 < \ga < 1:
X\sim Y \Longrightarrow \ga X+(1-\ga)Y \pp X.
\end{equation}
\noi Preferences are {\it AC convex\/}
if the above implication is imposed only for AC acts $X$, $Y$.
\end{definition}

 \noi Convexity of preference is a common assumption in consumer theory
(\citealp{MWG95}). It is also called
 quasiconvexity or, sometimes, quasiconcavity because it is equivalent to the quasiconcavity of any representing function. It reflects a preference for smoothing, diversification, and hedging in the models discussed next. It is remarkable that the same mathematical condition that captures the utility of commodity bundles in consumer theory also provides a characterization of risk aversion in subjective expected utility, as this section shows. It also captures ambiguity aversion in the currently most popular ambiguity models, as shown in the following section.

We next present an appealing implication of AC convexity in Savage's expected utility, where utility is not assumed to be affine in outcomes, and utility curvature captures different risk (or uncertainty) attitudes. To avoid triviality, we assume
 {\it non-degenerateness\/}, i.e., there exists an event $A$ with $0 < P(A) < 1$.

\begin{theorem} \label{Ttm.7:Savage.U.conc}
Assume Structural Assumption \ref{struct.ass.2} with
non-degenerate
expected utility and a 
mixture continuous utility function $U$.
The following three statements are equivalent.
 \begin{enumerate}[label=(\roman*)]
\item  $\succcurlyeq$ satisfies convexity.
\item  $\succcurlyeq$ satisfies AC convexity.
 \item $U$ is concave.
 \end{enumerate}
\end{theorem}

\cite{DK82} showed that (i) and (iii) in the theorem are equivalent, more generally, even without assuming continuity of utility, for Euclidean spaces instead of mixture spaces. We follow their proof closely, with some modifications to ensure AC. The main complication in the proof is that some convenient monotonicity properties in Euclidean spaces\footnote{For
 instance, we have no monotonicity of $U$ in $\ga$ in Eq.\ \ref{Hardy.et.al} in the proof given later.} are not available for general mixture spaces.

\citeauthor{WY19}'s (\citeyear{WY19}) Corollary 6 shows that the statements in Theorem \ref{Ttm.7:Savage.U.conc} are equivalent to comonotonic convexity of $\pp$. That is, in this case, the comonotonic and AC restrictions are equivalent. The characterization in Theorem \ref{Ttm.7:Savage.U.conc} and, similarly, in the related works just cited, through convexity of preference with respect to outcome mixing, is appealing because it makes risk aversion directly testable for subjective probabilities. To explain this point, we first note that concave utility captures risk aversion under expected utility. In decision under risk, where probabilities are objective and known beforehand, the conditions most commonly used to characterize risk aversion involve a preference for expected value or an aversion to
 mean-preserving spreads. Those conditions use probabilities as inputs. This use is problematic for decision under uncertainty because then probabilities are subjective and not directly observable, as in \cite{S54}. The main purpose of preference axiomatizations is to make theoretical properties directly observable. Therefore, the
 aforementioned common conditions for risk aversion, using probabilities as input, are not well suited for the context of uncertainty. Theorem \ref{Ttm.7:Savage.U.conc} and its predecessors make risk aversion directly observable and testable for subjective probabilities.

The contribution of our Theorem \ref{Ttm.7:Savage.U.conc} to its predecessor \cite{DK82} is, again, related to the central topic of this paper: we only need to inspect the most critical cases with maximal hedging possibilities. If risk aversion (and convexity, i.e., preference for diversification) passes those tests, then it holds everywhere.

\section{Anticomonotonic convexity for ambiguity: new properties}\label{Sect.new.props}

\noi This and the following section, like the preceding one, study AC restrictions for convexity. However, we now take a dual approach. Contrary to the preceding section, but as in all other sections, we assume linear/affine utility (e.g., EU) of outcomes. But now, unlike the preceding sections, we allow for nonlinear event weighting. That is, we investigate the implications of AC convexity for ambiguity models, deviating from EU for acts. Now, for the first time in this paper, new properties and models will result from the AC restriction. 

In \S \ref{class.add} and Theorem \ref{Ttm.4_charact_SEU_ind}, we presented a version of AA's setup for their axiomatization of expected utility. However, this setup has proved extremely useful for developing deviations from expected utility to capture ambiguity, and this section will use it. Famous contributions include \citeauthor{GS89}'s (\citeyear{GS89}) axiomatization of multiple priors and \citeauthor{S89}'s (\citeyear{S89}) axiomatization of Choquet expected utility, initiating the field of ambiguity theory. 

We present results for Schmeidler's CEU model.
A {\it weighting function\/} $W$ maps events to $[0,1]$ and satisfies $W(\ey) = 0$, $W(\gww) = 1$, and $A\sps B \Ra W(A) \geq W(B)$. We call $W$ {\it convex\/} if $W(A\cup B) + W(A\cap B) \geq W(A) + W(B)$ for all events. This implies
 {\it pseudo-convexity\/}: $W(A) \leq W(A\cup B) - W(B) \leq 1 - W(A^c)$ for all disjoint events $A,B$. {\it Choquet expected utility\/} ({\it CEU\/}) holds if there exists a weighting function $W$ and an affine utility function $U:\SC\ra \R$ such that the preference relation maximizes
\begin{equation}
X \mapsto \int_{[0,\iy)} W(U(X)\geq x) \d x
- \int_{(-\iy, 0]} (1 - W(U(X)\geq x)) \d x.
\end{equation}

We again study the convexity of preference. Any utility effect, as in Theorem \ref{Ttm.7:Savage.U.conc}, has now been ruled out by the affinity assumption of $U$. Hence, as follows from Theorem \ref{Ttm.7:Savage.U.conc}, convexities must now speak to deviations from EU. In the first axiomatized ambiguity models (\citealp{GS89}; \citealp{S89}), and in many that followed later, convexity was found to be equivalent to ambiguity aversion, explaining \citeauthor{E61}'s (\citeyear{E61}) famous paradox. Hence, convexity has as yet been the most central condition in ambiguity theories.

This section presents a case where an AC restriction essentially weakens a preference condition, i.e., convexity in the AA setup. \cite{ACV21} (their Theorem 1 and Corollary 1) first proved the following result for the special case of $\SC = \R$ and linear utility. Their result is deep, with a complex proof. We next provide its extension to general mixture spaces, which readily follows, thus covering AA's setup.

\begin{proposition}\label{Prop.8.AC=pseudo.mix.space} Assume Structural Assumption \ref{struct.ass.2} and CEU. Then AC convexity of $\pp$ is equivalent to
 pseudo-convexity of $W$.
\end{proposition}

Proposition \ref{Prop.8.AC=pseudo.mix.space} implies that AC convexity of $\pp$ is strictly more general than convexity because
 pseudo-convexity of $W$ is clearly more general than convexity, and the latter is equivalent to convexity of $\pp$ (\citeauthor{{S89}}, \citeyear{{S89}}). Example \ref{Ex.10.AC.conv.noncov.} below will confirm that AC convexity is strictly more general.

\section{Anticomonotonic convexity for ambiguity: new models}\label{Sect.new.models}

\noi We now turn to a case where the AC restriction brings a more general model. We first define the model. It is a subcase of Schmeidler's CEU. The
 {\it double-cautious\/} ambiguity model holds if $\pp$ maximizes CEU with respect to an affine utility function $U:\SC \ra \R$ and a weighting function $W$ that is 
{\it e(vent)-cautious\/:} $[W(E) > 0 \Ra W(E^c) = 0]$ and 
{\it w(eight)-cautious\/}: $W(E) \leq 0.5$ for all $E \neq \gww$. As for the intuition of these two conditions, the proof of Proposition \ref{Prop_9_DoubleC} shows that 
 e-cautiousness is equivalent to the next condition, clarifying its cautiousness interpretation: one is allowed to hope for something good ($\textnormal{CEU}(X)$ or more) only if it is very likely in the sense that getting less is
 quasi-impossible. Thus, what one hopes for is cautious in the sense that it can still qualify as a kind of 
worst-case scenario.
\begin{equation}\label{e.cautious}
\textnormal{For~all~acts~}X\textnormal{~and~} \xee > 0:
W\{\gw \elt \gww: U(X(\gw)) < \textnormal{CEU}(X) - \xee\} = 0.
\end{equation}
For the intuition of
 w-cautiousness, we define $\textnormal{IU}(X) := \inf_{\gw \in \gww} (U(X(\gw))$ for each act $X$. It is
 real-valued because acts are bounded.
The proof of Proposition \ref{Prop_9_DoubleC} shows that 
w-cautiousness is equivalent to the next condition, clarifying its cautiousness interpretation:
if one hopes for something good ($\xee$ more than the worst case), then its bad opposite (even if
 quasi-impossible), should still receive at least as much attention (decision  
\begin{equation}\label{w.cautious}
\textnormal{For~all~acts~}X\textnormal{~and~}
\xee > 0:
W\{\gw \elt \gww: U(X(\gw)) > \textnormal{IU}(X)+\xee\} \leq 0.5.
\end{equation}

We next turn to a preference axiomatization of the
 double-cautious model. For a preference axiomatization of CEU in the AA setup, \cite{S89} gave necessary and sufficient conditions, mainly comonotonic independence. They could be added in the theorem below to obtain a complete preference axiomatization, but for brevity we will not repeat them. By $x_E y$ we denote the
 two-outcome act that assigns outcome $x$ to event $E$ and $y$ to $E^c$. We say that $\pp$ satisfies
{\it e-cautiousness\/} if, for all outcomes $x \spp y$ and events $E$, $[x_{E}y \spp y \Ra y_{E}x \sim y]$.
We say that $\pp$ satisfies
{\it w-cautiousness\/} if, for all outcomes $x \spp y$ and events $E \neq \gww$, $x_{E}y \rp 0.5x + 0.5y$.

\begin{proposition}\label{Prop_9_DoubleC} 
Assume Structural Assumption \ref{struct.ass.2} and   CEU. The following three statements are equivalent.
\begin{enumerate}[label=(\roman*)]
    \item The double-cautious model holds.
    \item Conditions (\ref{e.cautious}) and (\ref{w.cautious}) hold.
    \item $\pp$ is e-cautious and w-cautious.
\end{enumerate}

Further, the double-cautious model satisfies AC convexity.
\end{proposition}

E-cautiousness and w-cautiousness only involve AC acts and, hence, the AC restriction is vacuous for these preference conditions. The convexity preference condition does involve acts that are not AC and here the AC restriction turns out to provide a real restriction in the premise, leading to a less restrictive preference condition. Thus, the
 double-cautious model implies AC convexity but not convexity, as the following example shows.

\begin{example}\label{Ex.10.AC.conv.noncov.} Let: $\gww = [0,1]$, $\F$ is the usual Lebesgue
 sigma-algebra, $\SC = \R^{+}$, and $U$ is the identity ($U(x) = x$ for all $x$). To define $W$, let $\gl$ be the usual Lebesgue measure (uniform distribution). Let $g:[0,1] \ra [0,1]$ be nondecreasing, $g(p) = 0$ for all $0 \leq p < 0.5$, $0 \leq g(p) \leq 0.5$ for all $0.5 \leq p < 1$, $g(1) = 1$. Further, $g$ is nonconvex on $[0.5, 1)$, say $g(p) = \sqrt{(2p -1)}/2$ there. We define $W(E) = g(\gl(E))$ with one exception: if $\gl(E) = 1$ but $E \neq \gww$, then $W(E) = 0.5$ rather than 1. This $W$ is double cautious so that $\pp$ is AC convex. Further, $W$ is not convex as readily follows from nonconvexity, even strict concavity, of $g$ on $[0.5,1)$, and, consequently (\citeauthor{{S89}}, \citeyear{{S89}}) neither is $\pp$.
 The latter claim is verified by calculations in Appendix B.
\end{example} 

In general, Example \ref{Ex.10.AC.conv.noncov.} but with $g(p) = (2p-1)^\theta/2$ on $(0.5,1)$ for some $\theta > 0$, gives a convenient parametric family for the 
 double-cautious model. Conditions (\ref{e.cautious}) and (\ref{w.cautious}) are conceptually simpler and easier to implement than convexity: they are directly imposed on the evaluation made of a relevant act $X$, rather than involving inspection of mixtures of acts. In this sense, the relaxation of convexity, maintaining AC convexity, is useful. Further, Example \ref{Ex.10.AC.conv.noncov.} suggests that the extra caution coming from convexity is not big.

The results presented in the last two sections primarily serve to demonstrate the possibility of getting new properties and models from AC.  Detailed studies of the pros and cons of such models and further models and properties to be derived from AC restrictions are left to future work. The end of Appendix A cites some results from the literature that may be useful for such future work.

\section{Conclusion}\label{Conclusion}

\noi This paper provides a systematic study of anticomonotonic restrictions of axioms for preference relations and functionals. Anticomonotonicity is the natural counterpart to the
 well-known comonotonicity. We obtained many generalizations of classical theorems, for each showing where the most critical tests are. These tests concern cases with maximal possibilities for hedging. Our results highlight the asymmetry between anticomonotonicity and comonotonicity. For ambiguity, anticomonotonicity can serve to bring new phenomena and models.

\vspace{\baselineskip}

\section*{Appendix A: Linear/affine functionals}

\noi This Appendix presents some results similar to Theorems \ref{Ttm.1.anticomon.lin.cont} and \ref{Ttm.4_charact_SEU_ind}, for linear/affine functionals.

The following lemma, repeating part of Theorem \ref{Ttm.1.anticomon.lin.cont} and used in its proof, is remarkable in giving, for finite state spaces, a complete logical equivalence of a condition and its AC restriction, i.e., (AC) additivity. We do not expect the equivalence to hold for general state spaces without some extra regularity condition, but this remains to us an open question. We maintain the notation $B(\gww,\F)$ below, although this set now contains all maps from $\gww$ to $\R$.

\begin{lemma}\label{Lemma.11_additive.Omega.finite}
Suppose that $\gww$ is finite and $\F=2^{\gww}$. For $I:B(\gww,\F) \to \R$, AC additivity is equivalent to additivity.
\end{lemma}

Theorem \ref{Ttm.1.anticomon.lin.cont}, using Lemma \ref{Lemma.11_additive.Omega.finite}, assumed a full linear space as domain and only used an elementary addition operation. The following Proposition considers more general convex sets as domain, involving convex combinations. It underlies Theorem \ref{Ttm.4_charact_SEU_ind}. For simplicity, and because we do not need more, we give it only for $\SC = \R$. For later reference, we repeat that, for $\SC = \R$ and a convex set $D$ of acts, $I:D \to \R$ is affine or linear if:
 \begin{equation}\label{def.aff.I}
\textnormal{For~all~} \ga\in \left[0,1\right] \textnormal{~and~} X,Y\in D:
I\left(\ga X+(1-\ga)Y\right)=\ga I(X)+(1-\ga)I(Y) .
\end{equation}

\noi We call $I$ {\it comonotonically affine\/} if Eq.\ \ref{def.aff.I} holds for $I$ whenever $X,Y$ are comonotonic. We call $I$ {\it AC affine\/} if Eq.\ \ref{def.aff.I} holds for $I$ whenever $X,Y$ are AC.

 \begin{proposition}\label{Prop.12.Anticomon.aff.cont.mon} Assume $\SC=\R$ and a functional $I : D \ra \R$, where $D\sbs B(\gww,\F)$ is convex and contains a constant act in its interior. Then AC affinity is equivalent to affinity whenever $I$ is monotonic (with respect to $\geq$ on $\R$) or continuous.
\end{proposition}

We now turn to an application to finance. Again $\SC = \R$, and now acts are financial assets and $I$ reflects the market price. Additivity of $I$ and even linearity are implied by common market trade assumptions and are thus automatically satisfied. Monotonicity is then taken as the critical condition in Proposition \ref{Prop.13_ttmnoarbitrage} below: a linear combination of trades should never lead to a sure loss
(no-arbitrage). Market prices $I$ are {\it normalized\/}: $I(0) = 0$ and $I(1) = 1$, implying, together with the other conditions, that $I(x) = x$ for all outcomes $x$.
The fundamental theorem of asset pricing entails that
 no-arbitrage implies
 as-if risk-neutral pricing: there exists a probability measure $P$ on $\gww$ such that $I$ is its expectation, denoted $\E_P$ or $\E$ for short. We generalize this fundamental theorem of finance. First, for linear combinations, only additivity is needed, and no scalar multiplication. (This point has been known for long time.) We further show that additivity can be weakened to AC acts. That is, the critical test of
 no-arbitrage in financial markets occurs in cases where leverage possibilities are maximal. This suffices to ensure
 no-arbitrage everywhere.

\begin{proposition}\label{Prop.13_ttmnoarbitrage} There exists a probability measure $P$ such that $I = \E_P$
(``as-if risk-neutrality'') if and only if $I$ is normalized and satisfies monotonicity and AC additivity. Here, $P$ is unique.
\end{proposition}
\noi
\cite{GS22} characterized
 no-arbitrage for arbitrary sets of acts and discussed its normative status.

In the risk management literature, for a risk measure $I$,
the equality $I(X+Y)=I(X)+I(Y)$ is often interpreted as that no diversification benefit\footnote{The
diversification benefit often refers to $I(X)+I(Y) -I(X+Y)$; see \cite{MFE15}.}
is assigned to the portfolio vector $(X,Y)$; see \cite{WZ21} in the context of the Basel Accords.
In this context, Proposition \ref{Prop.13_ttmnoarbitrage} is intuitive: If no portfolio of two AC risks (representing maximum hedging effect) is assigned a diversification benefit, then no portfolio should have any diversification benefit, and hence the risk measure should simply be the expected value.
This is in sharp contrast to the idea of assigning no diversification benefit to comonotonic risks, which leads to a large class of risk measures called distortion risk measures; mathematically, they coincide with the dual utility functionals of \cite{Y87}. See \cite{MFE15} for the use of distortion risk measures in risk management.

We next turn to de Finetti's book making argument. Again, $\SC = \R$.
 {\it Subjective expected value\/}, or {\it expected value\/} ({\it EV\/}), holds if EU holds with $U$ the identity function. {\it Additivity\/} holds for $\pp$ if 
 \begin{equation}\label{def.add.pr}
\textnormal{for~all~acts~} X,Y,Z:
X \sim Y \Longrightarrow X+Z \sim Y+Z.
\end{equation}
If a certainty equivalent exists for every act, as is the case in all results in this paper, then a convenient reformulation is:
 \begin{equation}\label{def.add.pr.x}
\textnormal{for~all~acts~} X,Z \textnormal{~and~outcomes~}x:
X \sim x \Longrightarrow X+Z \sim x+Z.
\end{equation}
The condition at first seems to be weaker than Eq.\ \ref{def.add.pr} because of the restriction to constant $Y=x$. However, it readily implies Eq.\ \ref{def.add.pr} by
 two-fold application with the (same) CE for $X$ and $Y$ and transitivity. The condition is
 well-suited for our purposes because the constant act $x$ is automatically AC with the other acts.

\begin{definition}
{\it AC additivity\/} holds for $\pp$ if Eq.\ \ref{def.add.pr.x} is imposed only if $X$ and $Z$ are AC.
\end{definition}

\begin{proposition}\label{Prop.15_ttmdeFinetti} Assume Structural Assumption \ref{struct.ass.2} with $\pp$ on $\SC = \R$ the natural ordering $\geq$.
There exists a probability measure $P$ such that expected value holds if and only if there exists a certainty equivalent for every act and weak ordering, monotonicity, and AC additivity hold.
\end{proposition}

 de Finetti and many other authors who have written about bookmaking assumed additivity more or less implicitly\footnote{Whereas
 this assumption is natural in finance, it is highly restrictive in the present context of individual choice. The bookmaking argument usually makes yet stronger assumptions by also incorporating positive scalar multiplications and, thus, positive linear combinations. Proposition \ref{Prop.15_ttmdeFinetti} showed that such assumptions are not needed because they are implied by the other conditions.},
 but emphasized the importance of monotonicity. They used the above result, without the AC restriction, and several variations, to argue that it is rational to use subjective probabilities in the context of uncertainty. Linearity of utility, as implied here, is reasonable for moderate stakes (\citealp{lHV19}, p.\ 189; \citealp{S54}, p.\ 91). de Finetti's result was historically important as a foundation of Bayesianism. Our result shows that the most critical case of bookmaking occurs when there are maximal possibilities of hedging (AC). That is, de Finetti needed to defend his condition only for AC cases.

Next, we suggest a generalization of AC, similar to the following generalization of comonotonicity that we explain first.\footnote{We
 thank a referee and editor for encouraging us to discuss possible further generalizations.}
 Two acts $X$ and $Y$ are {\it maxmin related\/} if for every state $\gw$ either $X$ takes its best value or $Y$ takes its worst value, or vice versa. This implies that $X,Y$ are comonotonic. Remarkably, many results in the literature using comonotonic preference conditions can be generalized by imposing the condition only for maxmin related acts. This way, and historically remarkable, \cite{A77} preceded \cite{S86} by providing a more general axiomatization of the Choquet integral. Other papers providing such maxmin generalizations of comonotonicity include \cite{C91}, \cite{ACV21}, \cite{W90}, \cite{BCC24}, and \cite{CMM15}, whose 
 put-call parity conditions are equivalent to \citeauthor{{A77}}'s maxmin relatedness. Even if mathematically more general than comonotonicity, maxmin relatedness never became very popular. We think that this happened because comonotonicity is more intuitive and better at capturing conceptual and empirical content. We can similarly generalize AC, e.g., if $\SC$ $=$ $\R$, by requiring $X,-Y$ to be maxmin related. That is, either (1) at every $\gw$, either $X$ or $Y$ is best, or (2) at every $\gw$, either $X$ or $Y$ is worst. We conjecture that the AC condition can be generalized in this manner in several results in our paper. We did not pursue this generalization because we find AC more intuitive, similarly as the literature has preferred comonotonicity to maxmin relatedness.

Finally, we briefly mention some results from the literature that may be useful for further studies of AC restrictions.
\cite{ACV21} provided many related results for superadditivity, supermodularity, and other properties, and implications for uncertainty attitudes and diversification.
\cite{BW22} provided optimization techniques for nonexpected utility models that are neither differentiable nor satisfy convexity of preference. Under some further assumptions, \cite{CCMTW22} axiomatized their
 star-shaped representing functionals through the following condition, weaker than AC convexity. {\it Uncertainty reduction\/} holds if:
\begin{equation}
\textnormal{For~all~acts~} X, \textnormal{~outcomes~}x, \textnormal{~and~}
0 < \ga < 1:
X \sim x \Longrightarrow \ga X + (1-\ga)x \pp X.
\end{equation}
 \noi The condition is weaker than AC convexity because every constant act $x$ is AC with every other act. Thus, AC convexity is between convexity and uncertainty reduction.
Given the other assumptions, AC convex functionals will thus be in the ``middle'' between convex and
 star-shaped functionals. Interestingly, \cite{CCMTW22} showed that their
 star-shaped functionals are maxima of concave functionals, a result that can be used to analyze AC convex preferences and functionals.

\section*{Appendix B: Proofs}

\noi We present proofs of results in the order of appearance in the main text and then in Appendix A. This is not a logical order in the sense that some proofs use results presented later. We then indicate those in the beginning of proofs.\vspace{\baselineskip}

\noi {\sc Proof of Theorem \ref{Ttm.1.anticomon.lin.cont}.}
This proof uses Lemma \ref{Lemma.11_additive.Omega.finite}. 

It is direct that linearity implies additivity, which implies AC additivity. We, therefore, assume the latter and derive linearity.

For any fixed finite partition, AC additivity implies additivity for the simple acts defined on that partition by Lemma \ref{Lemma.11_additive.Omega.finite}. Theorem 5.1.1 in \cite{AC66} shows that linearity follows for these acts under mild extra conditions such as continuity (at one point suffices) or monotonicity. Linearity follows for all simple acts because any pair of simple acts is measurable w.r.t.\ a joint simple partition. Finally, by standard integration techniques, linearity extends to all bounded acts: each can be ``sandwiched'' between dominating and dominated simple functions. By continuity or monotonicity, its $I$ value then is the limit of the $I$ values of the limiting simple acts. \ep
\vspace{\baselineskip}

\noi {\sc Proof of Theorem \ref{Ttm.4_charact_SEU_ind}.} This proof uses Proposition \ref{Prop.12.Anticomon.aff.cont.mon}.

That (iii) implies (i), and (i) implies (ii), is direct. We, therefore, assume (ii) and derive (iii). If all outcomes are indifferent then so are, by monotonicity, all acts and, hence, the result is trivial, with $U$ constant. So, we assume nontriviality. On the outcome set standard mixture independence axioms hold because AC does not impose any restriction. By \cite{HM53}, there exists an affine representation on outcomes. We, until further notice, fix two outcomes $M \spp m$, and consider only acts $X$ with $M \pp X(\gw) \pp m$ for all $\gw$. By monotonicity and mixture continuity, for each such act there exists a $0 \leq p \leq 1$ such that $pM + (1-p)m \sim X$. By Theorem 4 of \cite{HM53}, $p$ is uniquely determined and represents preferences over acts. We denote it by $\textnormal{MP}(X)$, the {\it matching probability\/} of $X$. It can be taken as a certainty equivalent for each act. We next show that MP is an expectation representation for all acts (by, essentially, establishing Cauchy's equation for it).

We write $p^{*} = pM + (1-p)m$ for all $p \in [0,1]$. The idea of the proof is to replace all outcomes by their equivalent $p^{*}$, which by monotonicity does not affect preference, and then by isomorphisms everything follows from preceding results. The switches between isomorphic spaces below involve some notational burden.

We first show that $\textnormal{MP}$ is AC affine. Assume $X$ and $Y$ AC and $\ga\in (0,1)$. Write $p = \textnormal{MP}(X)$ and  $q = \textnormal{MP}(Y)$.
Now
\[\ga X + (1-\ga)Y \sim \ga p^{*} + (1-\ga)Y \sim \ga p^{*} + (1-\ga) q^{*} =  (\ga p + (1-\ga) q)^{*},\]
 where the first two equivalences follow from AC independence and the last equality from affinity of MP on outcomes (also readily and more basically from distributivity in mixture spaces). The equality $$\textnormal{\textnormal{MP}}(\ga X + (1-\ga)Y) = \ga p + (1-\ga)q$$
 follows: $\textnormal{\textnormal{MP}}$ is AC affine.

To invoke Proposition \ref{Prop.12.Anticomon.aff.cont.mon}, we adjust the domain of $\textnormal{\textnormal{MP}}$ to become a subset of $B(\gww,\F)$.
For each act $X$, we define $X' : \gww \ra  [0,1]$ by $X'(\gw) = \textnormal{\textnormal{MP}}(X(\gw))$ for all $\gw$. This $X'$ is measurable because every inverse of a preference interval is an event, and $X'$ is also bounded. Define $I$ by $I(X') = \textnormal{\textnormal{MP}}(X)$. This $I$ is
well-defined because all $X$ with the same $X'$ are indifferent by monotonicity. This $I$ inherits monotonicity from $\textnormal{MP}$. It is also AC affine: Consider AC $X'$, $Y'$ and $0 <\ga < 1$. We take underlying $X,Y$ with $X(\gw) = X'(\gw)^{*}$ and $Y(\gw) = Y'(\gw)^{*}$; they are also AC. For every $\gw$,
\[(\ga X(\gw) + (1-\ga) Y(\gw))' = \ga X'(\gw) + (1-\ga) Y'(\gw)\]
 because  $\textnormal{MP}$ is affine on outcomes. Hence,
\[I(\ga X' + (1-\ga) Y') = \textnormal{MP}(\ga X + (1-\ga)Y).\]
 By AC affinity of $\textnormal{MP}$, this is
$\ga \textnormal{MP}(X)  + (1-\ga) \textnormal{MP}(Y) = \ga I(X') + (1-\ga)I(Y')$; $I$ is AC affine. It is affine by Proposition \ref{Prop.12.Anticomon.aff.cont.mon}.
 It is normalized.

By standard techniques (e.g., $I$'s affinity implies strong independence) $I$ is $\E_P$ for a probability measure $P$, first for all indicator functions, then for all simple $X'$, and then, by monotonicity, for all $X'$. Because $\textnormal{MP}(X) = I(X')$, $\textnormal{MP}$ is the EU functional with $\textnormal{MP}$ on outcomes as affine utility function $U$.
We have obtained the desired representation for all acts with outcomes between $m$ and $M$.

We now turn to acts with outcomes not between $m$ and $M$. For any other outcomes $M^{*} \pp M \pp m \pp m^{*}$ we can similarly obtain an expectation representation. We can rescale all these to take value 0 at $m$ and value 1 at $M$. They then all agree on common domain and are all part of one expectation functional defined on the whole domain. \ep
\vspace{\baselineskip}

\noi {\sc Proof of Theorem \ref{Ttm.7:Savage.U.conc}.}
It is clear that (iii) implies (i), and (i) implies (ii). We, therefore, assume (ii), and derive (iii). Assume, for contradiction, that $U$ is not concave. Then there are outcomes $M', m'$ and $0 < \ gl' < 1$ such that
\[U(\ga' M' + (1-\ga')m') < \ga' U(M') + (1-\ga')U(m').\]
 By mixture continuity, we can find the largest $0 \leq \gs < \ga'$ such that $m = \gs M' + (1-\gs)m'$ satisfies
$U(m) = \gs U(M') + (1-\gs)U(m')$ and the smallest $1 \geq \gt> \ga'$ such that $M = \gt M' + (1-\gt)m'$ satisfies  $U(M) = \gt U(M') + (1-\gt)U(m')$.
 We have
\begin{equation}\label{Hardy.et.al}
 \textnormal{~for all~} 0 < \ga < 1: U(\ga M + (1-\ga)m) < \ga U(M) + (1-\ga)U(m).
\end{equation}

By
 non-degenerateness, we can take $A\in \mathcal F$ with $0 < P(A) = p < 1$.  We write $(x,y)$ for $x_A y$ and in the rest of this proof use only such acts. First assume $(M,m) \sim (m,M)$. This occurs if $P(A) = 0.5$ or $U(m) = U(M)$. Note that the two acts are AC. By AC convexity,
\[((m+M)/2, (m+M)/2) \pp (m,M),\]
 implying
\[U((m+M)/2) \geq (U(m) + U(M))/2,\]
 contradicting Eq.\ \ref{Hardy.et.al}. From now on we may assume $U(M) >  U(m)$ and $p = P(A) > 0.5$. Otherwise we would interchange $M$ and $m$, and/or $A$ and $A^c$. We have $(M,m) \spp (m,M)$. In the remainder of this proof, we will only use outcomes $x$ of the form $x = \ga M + (1-\ga) m$ for some $\ga$. We assume without further mention that al outcomes are of this form. Mapping $\ga$ to $\ga(x)$ provides an isomorphism of interval ${[}0,1]$ to the outcome space.\footnote{In
general, the only way in which the mixture space of all $x = \ga M + (1-\ga) m$ may not be isomorphic to $[0,1]$ is by having $x = \ga M + (1-\ga) m = \ga' M + (1-\ga') m$ for all $0 < \ga < \ga' < 1$, as follows mainly from distributivity. This case is excluded by Eq.\ \ref{uxj} below.}
 We use it for defining average increases below.

 We define $x_0 =m $. By mixture continuity, there exists $m \srp x_1 \srp M$ with $(x_1,m) \sim (x_0,M)$. If there are several such, we take the one closest to $m$, i.e., we take $x_1 = \ga M + (1-\ga)m$ with $\ga$ minimal (existing by continuity of $U$).
 By mixture continuity we can inductively define a ``standard sequence''
$m=x_0, x_1, x_2, \ldots, x_n$ such that
$(x_{j+1},m) \sim (x_j,M)$ for all $j < n$ each $x_j$ closest to $m$ so that
 $\ga(x_{j+1}) > \ga(x_j)$
 and
$(M,m) \srp (x_{n},M)$. We have
\begin{equation}\label{uxj}
 \textnormal{~for all~}  j: U(x_{j+1}) - U(x_j) = \frac{(1-p)(U(M) -U(m))}{p}.
\end{equation}
 We first consider the case $x_n \srp M$. We then similarly define a ``standard sequence''
$M=y_{n+1}, y_n, y_{n-1} \ldots, y_1$ such that
$(y_{j-1},M) \sim (y_j,m)$
and
 $\ga(y_{j-1}) < \ga(x_{j-1}) < \ga(y_j)$ and $y_j$ closest to $m$
 for all $j$. We have $(m,M) \spp (y_{1},m)$ and
\begin{equation}\label{uyj}
 \textnormal{~for all~}  j: U(y_{j}) - U(y_{j-1}) = \frac{(1-p)(U(M) -U(m))}{p}.
\end{equation}

For every $j$ we have $x_{j-1} \srp y_j \srp x_j$
and, further, there exists a $0 < \ga < 1$, dependent on $j$, such that
$y_j = \ga x_j  + (1-\ga)x_{j-1}$.
By $m \rp x_{j-1} \rp x_j \srp M$, we have AC of  $(x_{j-1},M)$ and $(x_j,m)$. AC convexity and
$(x_{j-1},M) \sim (x_j,m)$ imply
\begin{equation}\label{triplepref}
\ga (x_j,m) + (1-\ga) (x_{j-1},M)
\pp
(x_j,m) \sim (x_{j-1},M).
\end{equation}
\noi We next show that, because the triple of outcomes $m$, $\ga M + (1-\ga)m$, $M$ on $A^c$ in Eq.\ \ref{triplepref} bring in a kind of strict convexity, the triple $x_{j-1}$, $y_j (= \ga x_j + (1-\ga) x_{j-1})$, $x_j$ on $A$ must bring in a kind of concavity, and enough so to maintain the aforementioned AC convexity. This point is elaborated on next.

The $U$ value of the left act in Eq.\ \ref{triplepref} exceeds the $U$ value of the other two acts and, therefore, also the $\ga$/$1-\ga$
convex combination of the latter two $U$ values. That is,
\begin{align*}
&p U(\ga x_j  + (1-\ga) x_{j-1})  +
(1-p) U(\ga m  + (1-\ga) M)
 \\ &\geq  p   (\ga U(x_j)  + (1-\ga)U(x_{j-1}))
+ (1-p)   (\ga U(m) + (1-\ga)U(M)).
\end{align*}
 This and
\[(1-p) U(\ga m  + (1-\ga) M)  < (1-p)   (\ga U(m) + (1-\ga)U(M))\]
 (implied by Eq.\ \ref{Hardy.et.al})
imply (dropping $p$)
\[U(\ga x_j  + (1-\ga) x_{j-1})  >   \ga U(x_j)  + (1-\ga)U(x_{j-1}).\]
 The triple $x_{j-1}$, $y_j$ (which equals $\ga x_j + (1-\ga)x_{j-1}$), and $x_j$ exhibit a kind of concavity.

Using the above isomorphism with ${[}0,1]$, the aforementioned ``concavity''  means that the average increase of $U$ over ${[}x_{j-1},y_j]$ exceeds that over ${[}y_j, x_j]$:
\[(U(y_j) - U(x_{j-1}))\ga > (U(x_j) - U(y_j))/(1-\ga).\]

A similar proof shows that the average increase of $U$ over ${[}y_j, x_j]$ exceeds that over ${[}x_j,y_{j+1}]$. In this proof, write $x_j = \ga' y_j + (1-\ga')y_{j+1}$ and proceed as above with $y_j$ for $x_{j-1}$, $x_j$ for $y_j$, $y_{j+1}$ for $x_j$,
Eq.\ \ref{uyj} for Eq.\ \ref{uxj}, and $\ga'$ for $\ga$. The two results together imply that the average increase over an interval decreases as we move from $m$ to $M$ from $[y_j,x_j]$ to $[x_j, y_{j+1}]$, to $[y_{j+1},x_{j+1}]$, and so on.

By Eq.\ \ref{Hardy.et.al}, the average $U$ increase over ${[}m,y_1]$ is strictly below that of ${[}m,M]$. But we have, just, partitioned that interval ${[}m,M]$ into $2n+1$ intervals that all have a strictly smaller average increase than ${[}m,y_1]$. A contradiction has resulted.

We, finally, turn to the case of $x_n \sim M$. We take $z_0, \ldots, z_{2n}$ such that $z_j = \ga_j M + (1-\ga_j)m$, $z_{2j} = x_j$, $U(z_{2j+1}) =  (U(x_j) + U(x_{j+1}))/2$, $z_j$ closest to $m$, $\ga_{j+1} > \ga_j$ for all $j$.
 We also define $m' = \ga M + (1-\ga)m$ such that $U(m') = (U(M) + U(m))/2$.
By Eq.\ \ref{Hardy.et.al}, $\ga > 0.5$. We have $(z_j,M) \sim (z_{j+1},m') \sim (z_{j+2},m)$ for all $j$. By AC convexity, $\ga (z_j,M) + (1-\ga) (z_{j+2},m) \pp (z_j,M)$ and, hence,
 $\ga (z_j,M) + (1-\ga) (z_{j+2},m) \pp (z_{j+1},m')$. Hence,
$U(\ga z_j  + (1-\ga) (z_{j+2}) \geq (U(z_j)  + U(z_{j+2}))/2$ whereas $\ga < 0.5$. Given that $z_{j+1}$ was chosen closest to $m$, $\ga_{j+1}  < 0.5 \ga_j + 0.5 \ga_{j+2}$. The average increase of $U$ over $[z_j,z_{j+1}]$ strictly exceeds that over $[z_{j+1},z_{j+2}]$. This holds for all $j$. It is in contradiction with the average increase of $U$ over $[z_0, z_1]$ strictly being below that over $[m,M]$ (remember: $z_0 = m$) as follows from Eq.\ \ref{Hardy.et.al}.  \ep
\vspace{\baselineskip}

\noi {\sc Proof of Proposition \ref{Prop.8.AC=pseudo.mix.space}.} In words, we replace all outcomes by their $U$ values, extend $U(\SC)$ to all of $\R$ using positive homogeneity of the Choquet integral, and then use \cite{ACV21}.

We may assume that 0 is in the interior of the range of $U$, by setting $U(x) > 0 > U(y)$ for some outcomes $x \spp y$. Define $\gww' = \gww$, $\SC' = \R$, and $\SF'$ is the set of all measurable bounded maps from $\gww'$ to $\SC'$. Define $U'$ on $\SF'$ as the identity function, and let $\pp'$ on $\SF'$ maximize $\textnormal{CEU}'$ w.r.t.\ $U'$ and $W' = W$. Then for all acts $X,Y$ and new acts $X',Y'$ with $U(X(\gw)) = X'(\gw)$ and $U(Y(\gw)) = Y'(\gw)$ for all $\gw$, we have $X \pp Y \Lra X' \pp Y'$. In this way the new structure agrees with the original one and extends it. By positive homogeneity of $\textnormal{CEU}'$, the new structure satisfies AC convexity if and only if it does in a neighborhood around the constant new act 0.\footnote{ Multiply
 any pair of acts by $\ga > 0$ small enough to take them into that small neighborhood and verify AC convexity there.}
 That is, if and only if the original structure does. Our Proposition now follows from  \cite{ACV21}. \ep
\vspace{\baselineskip}

\noi {\sc Proof of Proposition \ref{Prop_9_DoubleC}.}
Eq.\ \ref{e.cautious} implies  
e-cautiousness of $W$: Assume $W(E) > 0$. Take outcomes $x \spp y$ and the act $X = x_E y$.
Then $\textnormal{CEU}(X) > U(y)$. Take $0 < \xee < \textnormal{CEU}(X) - U(y)$.
 By Eq.\ \ref {e.cautious}, $W(E^c) = 0$, as required by 
e-cautiousness of $W$.

E-cautiousness of $W$ implies
Eq.\ \ref{e.cautious}:
Assume, for contradiction, $ W(U(X) < \textnormal{CEU}(X) - \xee) > 0$ for some $\xee > 0$. Then
 e-cautiousness of $W$ implies $W(U(X) \geq \textnormal{CEU}(X) - \xee) = 0$, giving the contradiction
$ \textnormal{CEU}(X) \leq \textnormal{CEU}(X) - \xee$.

We have shown that Eq.\ \ref{e.cautious} is equivalent to
e-cautiousness of $W$, which is trivially equivalent to
e-cautiousness of $\pp$.

Eq.\ \ref{w.cautious} implies w-cautiousness of $W$:
Assume $E \neq \gww$. Take outcomes $x \spp y$, $X = x_E y$, and $0 < \xee < U(x) - U(y)$. By Eq.\ \ref{w.cautious}, with $\textnormal{IU}(X) = U(y)$,
 $W(E) \leq 0.5$. $W$ is
 w-cautiousness.

W-cautiousness of $W$ implies 
Eq.\ \ref{w.cautious}: for $\xee > 0$,
$\{\gw \elt \gww: U(X(\gw)) > \textnormal{IU}(X)+\xee\} \neq \gww$ so that, by
 W-cautiousness, its $W$ value does not exceed $0.5$. We have proved Eq.\ \ref{w.cautious}.

Eq.\ \ref{w.cautious} is equivalent to
w-cautiousness of $W$, which is trivially equivalent to
w-cautiousness of $\pp$.

We have shown equivalence of Statements (i), (ii), and (iii) without the AC convexity claim. We finally show that AC convexity can be added to Statement (iii).

\begin{lemma}\label{Lemma.16.AC.convex.DC} The
 double-cautious model satisfies AC convexity.
\end{lemma}

\noi {\sc Proof.}
By Proposition \ref{Prop.8.AC=pseudo.mix.space}, it suffices to derive
 pseudo-convexity of $W$. Assume disjoint events $A,B$, nonempty to avoid triviality. If $A\cup B = \gww$ then
$W(A\cup B) - W(B) \leq 1 - W(A^c)$ follows trivially. Otherwise, it follows from double cautiousness because 0.5 then is in between. We, finally, derive $W(A) \leq W(A\cup B) - W(B)$. It is trivial if $W(A) = 0$ and, hence, assume $W(A) > 0$. Then, by double cautiousness, $W(B) = 0$. We have
$W(A\cup B) - W(B) = W(A\cup B) \geq W(A)$ and we are done.  $QED$
\vspace{\baselineskip}

The proof of Proposition \ref{Prop_9_DoubleC} is done. \ep
\vspace{\baselineskip}

\noi {\sc Proof of nonconvexity in Example \ref{Ex.10.AC.conv.noncov.}.} 
First, $W$ violates convexity: take $A = [0, 0.58), B = [0, 0.5)\cup [0.58,0.66)$. Then
$W(A\cup B) + W(A\cap B) = \sqrt{0.08} + 0 < W(A) + W(B) =
\sqrt{0.04} + \sqrt{0.04} = 0.4$, violating convexity.
Further, $\pp$ also violates convexity:
Assume $X(\gw) = Y(\gw) = 2$ for all $\gw < 0.5$,
$X(\gw) = 1$ for all $0.50 \leq \gw < 0.58$, $X(\gw) = 0$ for all $\gw \geq 0.58$,
$Y(\gw) = 0$ for all $0.50 \leq \gw < 0.58$, $Y(\gw) = 1$ for all $0.58 \leq \gw < 0.66$, $Y(\gw) = 0$ for all $\gw \geq 0.66$.
Now $\textnormal{CEU}(X) = \textnormal{CEU}(Y) = (g(0.50) - g(0)) \times 2 + (g(0.58) - g(0.50)) \times 1 + 0 = 0 + 0.2 + 0 = 0.2$.
However, $\textnormal{CEU}((X+Y)/2) = (g(0.66) - g(0.50)) \times 1/2 = \sqrt{0.08}/2 = \sqrt{0.02} < 0.2$. Hence, $X \sim Y \spp (X+Y)/2$, violating convexity of $\pp$.  
\ep
\vspace{\baselineskip}

\noi {\sc Proof of Lemma \ref{Lemma.11_additive.Omega.finite}.}
We assume AC additivity and derive additivity.
Write $\gww =\{\gw_1,\dots,\gw_n\}$.

 \noi {\sc Step 1} [Additivity for $X$ and its AC $-X$]\\
Because $X$ and $-X$ are AC,
\[ I(0) = I(0+0) = I(0)+I(0) = 0\]
 and
\[0 = I(0) = I(X-X) = I(X)+I(-X),\]
 implying
 $I(-X)= -I(X).$

\noi {\sc Step 2} [Comonotonic additivity for $X$ and $Y$ from AC additivity for $X+Y$ and $-Y$]\\
For any comonotonic $X$ and $Y$, $X+Y$ and $Y$ are comonotonic so that $X+Y$ and $-Y$ are AC. Hence,
\[I(X)  =  I(X+Y-Y)  =  I(X+Y) + I(-Y)  =  I(X+Y) - I(Y).\]
 Comonotonic additivity follows.

\noi {\sc Step 3} [Additivity for general $X$ and $Y$ by writing as sums of increasing and decreasing functions and then comonotonic and AC additivity pairwise]\\
Consider two general $X,Y$.
With $\gww = \{\gw_1, \ldots, \gw_n\}$, we can write $X = X^\uparrow + X^\downarrow$ with $X^\uparrow(\gw_i)$ weakly increasing and $X^\downarrow(\gw_i)$ weakly decreasing in $i$, and $Y = Y^\uparrow + Y^\downarrow$ similar.
 By comonotonic additivity (CA) and AC additivity (ACA):
\begin{align*}
I(X+Y) &\xlongequal{(\rm def)} I(X^\uparrow + X^\downarrow+ Y^\uparrow + Y^\downarrow )
\\ & \xlongequal{(\rm ACA)}
I(X^\uparrow+ Y^\uparrow) + I(X^\downarrow+  Y^\downarrow )
\\&  \xlongequal{(\rm CA)}
I(X^\uparrow)+ I( Y^\uparrow) + I(X^\downarrow)+I(  Y^\downarrow )
\\& \xlongequal{(\rm ACA)}
I(X^\uparrow +X^\downarrow ) +I(  Y^\uparrow +   Y^\downarrow )  = I(X)+I(Y).
\end{align*}
 This shows that $I$ is additive. \ep
\vspace{\baselineskip}

\noi {\sc Proof of Proposition \ref{Prop.12.Anticomon.aff.cont.mon}.}
This proof uses Theorem \ref{Ttm.1.anticomon.lin.cont}.

We assume AC affinity and derive affinity under continuity or monotonicity. The reversed implication is trivial.

 We may assume $0 \in \textnormal{int}D$ and $I(0)=0$. To see this point, take a constant $k \elt \textnormal{int}D$. Define $D' = D-k$: $D'$ contains all acts resulting from subtracting $k$ from acts in $D$. Next define $I'$ on $D'$ correspondingly:
 $I'(X)  = I(X+k) - I(k)$. These $I'$ and $D'$ share all relevant properties, including AC, with $I$ and $D$. It suffices to prove our results for $I'$ and $D'$. We may omit primes.

The functional $I$ is positively homogeneous: For each $0 < \ga < 1$ and $X\in D$,
\[I(\ga X)  = I(\ga X + (1-\ga)0) = \ga I(X) + (1-\ga)I(0) = \ga I(X),\]
 using AC of $X$ and 0.

We next extend $I$ to $I^*$ defined on the whole vector space $B(\gww,\F)$ using positive homogeneity. That is, for each $X\in B(\gww,\F)$ we can find $\ga > 0$ so small that  $\ga X \in D$, and then define $I^*(X) = I(\ga X)/ \ga$. By associativity of scalar multiplication, $I^*$ is
 well-defined (independent of the particular $\ga$ chosen) and positively homogeneous.  Further, $I^*$ is AC affine because AC and AC affinity are compatible with multiplication by a common scalar. We next derive AC additivity of $I^*$. Consider AC $X, Y \in B(\gww,\F)$. Using positive homogeneity:
\[I^*(X+Y) = 2I^*(X/2 + Y/2) = 2 (I^*(X)/2 + I^*(Y)/2) = I^*(X) + I^*(Y).\] AC additivity holds for $I^*$.

Continuity of $I$ on $D$ implies continuity of $I^*$ on $B(\gww,\F)$, and monotonicity of $I$ similarly extends to $I^*$. Hence, under continuity, $I^*$ is linear by Theorem \ref{Ttm.1.anticomon.lin.cont}. Under monotonicity, $I^*$ is linear by Proposition \ref{Prop.13_ttmnoarbitrage} applied to the normalization of $I^*$ (dividing it by $I^*(1)$). Affinity of $I^*$ and $I$ follows. \ep
\vspace{\baselineskip}

\noi {\sc Proof of Proposition \ref{Prop.13_ttmnoarbitrage}.}
This proof uses Theorem \ref{Ttm.1.anticomon.lin.cont}.

It is direct that $I = \E_P$ implies the conditions of $I$. We, therefore, assume those conditions and derive
$I = \E_P$.

For any fixed finite partition, AC additivity implies linearity for the simple acts defined on that partition. This follows from Theorem \ref{Ttm.1.anticomon.lin.cont}.
  Linearity follows for all simple acts because any pair of simple acts is measurable w.r.t.\ a joint finite partition. To obtain the $\E_P$ representation for all simple acts, define $P(E) = I(1_E)$ for all $E$, which is nonnegative by monotonicity. It uniquely determines $P$.
We have $P(\gww) = 1$ because $I$ is normalized. Linearity implies additivity of $P$, and $I = \E_P$ for all simple functions.

Next, by standard integration techniques, the expectation is extended to all bounded acts: Each can be ``sandwiched'' between dominating and dominated simple acts. Its $I$ value is the limit of the $I$ values of the limiting simple acts, that is, $\E_P$, as we show in the remainder of this proof. For some $\xee > 0$ and simple acts $X$ and $Y$, assume $|X(\gw) - Y(\gw)| \leq \xee$ for all $\gw$. Then
\[|I(X) - I(Y)| = |I(X-Y)| \leq I(|X-Y|)  \leq  I(\xee)\]
 by monotonicity, and the latter tends to 0 for $\xee$ tending to 0 by linearity of $I$ on simple (including constant) acts. \ep
\vspace{\baselineskip}

\noi {\sc Proof of Proposition \ref{Prop.15_ttmdeFinetti}.}
This proof uses Proposition \ref{Prop.13_ttmnoarbitrage}.

EV directly implies the other conditions. We next assume the other conditions and derive EV. To derive AC additivity of the certainty equivalent (CE) functional (uniquely defined given that $\succcurlyeq$ coincides with $\geq$ on outcomes), assume $X,Y$ AC. Then $X \sim \textnormal{CE}(X)$ implies $X+Y \sim \textnormal{CE}(X) + Y$ and $Y \sim \textnormal{CE}(Y)$ implies $Y + \textnormal{CE}(X) \sim \textnormal{CE}(Y) + \textnormal{CE}(X)$. By transitivity, $X+Y \sim \textnormal{CE}(X) + \textnormal{CE}(Y)$. Thus, CE is AC additive. Further, it is monotonic and normalized. By Proposition \ref{Prop.13_ttmnoarbitrage}, it is $\E_P$. It represents $\pp$.  \ep
\vspace{\baselineskip}

\end{document}